\documentclass{article}

\usepackage{arxiv}

\usepackage[utf8]{inputenc} % allow utf-8 input
\usepackage[T1]{fontenc}    % use 8-bit T1 fonts
\usepackage{hyperref}       % hyperlinks
\usepackage{url}            % simple URL typesetting
\usepackage{booktabs}       % professional-quality tables
\usepackage{amsfonts}       % blackboard math symbols
\usepackage{nicefrac}       % compact symbols for 1/2, etc.
\usepackage{microtype}      % microtypography
\usepackage{lipsum}
\usepackage{graphicx}
\usepackage{amsmath}
\usepackage{multirow}
\usepackage{hhline}
\usepackage{tabularray}
\usepackage{booktabs, caption, makecell}
\usepackage{threeparttable}
\usepackage{cite}
\graphicspath{ {./images/} }
\newcommand{\minitab}[2][l]{\begin{tabular}{#1}#2\end{tabular}}

\title{MC BTS: simultaneously resolving magnetization transfer effect and relaxation for multiple components}

\author{
 Albert Jang \\
  Athinoula A. Martinos Center for Biomedical Imaging\\
  Massachusetts General Hospital\\
  Charlestown, MA 02129 \\
  Harvard Medical School\\
  Boston, MA 02114 \\
  \texttt{awjang@mgh.harvard.edu} \\
  \And
 Hyungseok Jang \\
  University of California, Davis\\
  Davis, CA 95616 \\
  \texttt{hyjang@health.ucdavis.edu} \\
  \And
 Nian Wang \\
  Advanced Imaging Research Center\\
  University of Texas Southwestern Medical Center\\
  Dallas, TX 75390 \\
  \texttt{Nian.Wang@UTSouthwestern.edu} \\
  \And
 Alexey Samsonov \\
  University of Wisconsin, Madison\\
  Madison, WI 53706 \\
  \texttt{samsonov@wisc.edu} \\
  \And
 Fang Liu \\
  Athinoula A. Martinos Center for Biomedical Imaging\\
  Massachusetts General Hospital\\
  Charlestown, MA 02129 \\
  Harvard Medical School\\
  Boston, MA 02114 \\
  \texttt{FLIU12@mgh.harvard.edu} \\
}

\begin{document}
\maketitle
\begin{abstract}
\textbf{Purpose:} To propose a signal acquisition and modeling framework for multi-component tissue quantification that encompasses transmit field inhomogeneity, multi-component relaxation and magnetization transfer (MT) effects.\newline
\textbf{Theory and Methods:} By applying off-resonance irradiation between excitation and acquisition within an RF-spoiled gradient-echo scheme, in combination with multiple echo-time acquisitions, both Bloch-Siegert shift and magnetization transfer effects are simultaneously induced while relaxation and spin exchange processes occur concurrently. The spin dynamics are modeled using a three-pool framework, from which an analytical signal equation is derived and validated through numerical Bloch simulations. Monte Carlo simulations were further performed to analyze and compare the model’s performance. Finally, the feasibility of this novel approach was investigated in vivo in human brain and knee tissues.\newline
\textbf{Results:} Simulation results showed excellent agreement with the derived analytical signal equation across a wide range of flip angles and echo times. Monte Carlo analyses further validated that the three-pool parameter estimation pipeline performed robustly over various signal-to-noise ratio conditions. Multi-parameter fitting results from in vivo brain and knee studies yielded values consistent with previously reported literature. Collectively, these findings confirm that the proposed method can reliably characterize multi-component tissue parameters in macromolecule-rich environments while effectively compensating for $B_1^+$ inhomogeneity.\newline
\textbf{Conclusion:} A signal acquisition and modeling framework for multi-component tissue quantification that accounts for magnetization transfer effects and $B_1^+$ inhomogeneity has been developed and validated. Both simulation and experimental results confirmed the robustness of this method and its applicability to various tissue types in the brain and knee.\newline
\newline
Keywords (3 to 6 words): Bloch-Siegert, magnetization transfer, BTS, multi-component, quantitative imaging
\end{abstract}

% keywords can be removed

\section{INTRODUCTION}
Quantitative MRI (qMRI) provides a powerful opportunity to elucidate detailed information regarding tissue microstructure and microenvironment. Compared to conventional qualitative MRI, quantitative techniques offer superior sensitivity in detecting pathologies and greater specificity for identifying disease subtypes. qMRI leverages tissue models grounded in spin physics to estimate biophysical parameters of interest. These parameters, including but not limited to relaxation, diffusion, perfusion, susceptibility and macromolecular content, have demonstrated sensitivity to a range of neurological and musculoskeletal conditions\cite{Seiler2021,Atkinson2019,Goveas2015,Raya2024,Huang2023,deVries2020,Harada2022,Wei2019,Harrison2015,Moazamian2025}. Specifically, in the brain, myelin, a bilayer lipid-rich cell essential for ensuring bioelectric signal conduction, forms an insulating sheath around the axon. Within its concentric layers, trapped water (i.e., myelin water) exhibits distinct properties compared to non-trapped water located inside axons and the surrounding extracellular space (i.e., axonal/extracellular water)\cite{Laule2007}. In the joint, articular cartilage is characterized by an extracellular matrix that provides structural integrity and mechanical function. The extracellular matrix is composed of water, with small amounts of macromolecules such as type II collagen and proteoglycans\cite{Crema2011}. Among these constituents, water that is tightly bound to macromolecules has unique spin relaxation properties compared to water that is loosely associated with matrix molecules\cite{Liu2015,Reiter2011,Reiter2011a,Reiter2009}.\newline

Due to the heterogeneous nature of biological tissues, multiparametric models are better suited to accurately characterize tissue. In particular, quantitative magnetization transfer (qMT) assesses the magnetization exchange between motion-restricted macromolecules and surrounding water protons. The widely used binary spin-bath model describes this exchange by modeling spin transfer between an aqueous free water pool and a restricted proton pool, providing insight into macromolecular content and integrity\cite{Sled2018}. Studies have applied qMT parameters to evaluate myelin sheath integrity in brain tissue\cite{Sled2018,Levesque2010,Kitzler2012} and collagen network breakdown in articular cartilage\cite{Moazamian2025,Stikov2011,Sritanyaratana2014}. For example, the decrease of macromolecular proton fraction derived from qMT has been found to correlate with the myelin degradation in multiple sclerosis\cite{Yarnykh2004,York2022}. Multi-component relaxometry is another quantitative technique that identifies multiple aqueous water pools and their relative fractions within tissue\cite{Menon1991}. Unlike qMT, multi-component relaxometry focuses on assessing relaxation features of the different aqueous water components, which reflect different water mobilities constrained by the mesoscale tissue architecture, for example, distinguishing myelin water from axonal/extracellular water in the brain\cite{Mackay1994}, or macromolecule-bound water from bulk water in the cartilage extracellular matrix\cite{Liu2015,Reiter2011,Reiter2011a,Reiter2009}. Specifically, many studies have found that the reduction of myelin water fraction can correlate with demyelination in neurodegenerative diseases such as dementia\cite{Faulkner2024} and multiple sclerosis\cite{Laule2018}, and the reduction of macromolecule-bound water fraction can reflect cartilage damage in osteoarthritis\cite{Liu2015,Liu2014}. More importantly, combining qMT and multi-component relaxometry may facilitate a better understanding of disease pathogenesis and progression. For example, a previous study demonstrated the value of integrating qMT and myelin water fraction measurements to investigate remyelination in longitudinal multiple sclerosis lesions\cite{Levesque2010}.\newline

In terms of imaging implementation, qMT can be conducted using gradient-echo pulsed irradiation schemes\cite{Sled2001}. Recent improvements of the pulsed model have included global fitting to correct biased parameter estimation\cite{Mossahebi2014}, dual-offset saturation strategies to compensate for on-resonance saturation and dipolar effects\cite{Soustelle2023}. On-resonance MT sensitivity\cite{Gloor2008} and selective inversion recovery techniques\cite{Gochberg2007} have also been used for quantifying MT parameters. Although these techniques have shown great potential in assessing qMT parameters, they are not without challenges. For example, a common practice in the pulsed irradiation model assumes the $T_1$ of the macromolecule pool is either equal to the free pool\cite{Mossahebi2014,Gloor2008,Yarnykh2002} or approximated as 1 s\cite{Yarnykh2004,Sled2001,Jang2023} for simplifying model fitting. However, these assumptions may cause bias in the actual values of the binary spin-bath parameters\cite{Helms2009,vanGelderen2016,Manning2021}. By nulling the longitudinal magnetization of the macromolecule pool in the acquisition process, selective inversion recovery methods do not rely on the direct knowledge of its $T_1$ value. However, due to its lack of multi-slice acquisition capability, it has limited volume coverage and can result in long scan times\cite{Gochberg2007}. Another common confounding factor is the transmit field ($B_1^+$) inhomogeneity, which leads to spatial variation of the actual flip angle applied within the imaging volume, and if not corrected, can cause significant qMT parameter estimation bias\cite{Mossahebi2014,Jang2023}.\newline

A common implementation of multi-component relaxometry uses exponential decay data from the multi-echo spin-echo sequences, such as the Carr-Purcell-Meiboom-Gill (CPMG)\cite{Mackay1994}, to estimate the distribution of $T_2$ values. More recently, efficient multi-echo spin-echo variants such as GRadient And Spin Echo (GRASE)\cite{Prasloski2012} have been employed for $T_2$-based methods. However, the repetition times used in these sequences, ranging from 1$\sim$2 s, are not sufficient for full longitudinal recovery, which, coupled with $B_1^+$ inhomogeneity, can introduce bias to the estimation process\cite{Prasloski2012a}. Similarly, $T_2^*$-based approaches from multi-echo gradient-echo sequences can also be used in multi-component relaxometry\cite{Du2007,Kijowski2017,Liu2017}. However, both $T_2$- and $T_2^*$- based approaches can be significantly affected by the MT effect\cite{Vavasour2000,Sati2013}. In the presence of macromolecules, there is a constant exchange of magnetization between free water protons and macromolecular protons, which unavoidably affects the spin dynamics of multiple free water components, resulting in both relaxation and MT-modulated MR signals. Studies have shown that these multiple components exhibit different magnetization transfer ratios\cite{Vavasour2000,Sati2013} and can cause significant errors in multi-component quantification when not properly accounted for\cite{Liu2016a}.\newline

Recently, we developed a qMT method called BTS (Bloch-Siegert and magnetization Transfer Simultaneously)\cite{Jang2023}. In BTS, a two-pool MT model consisting of an aqueous free water pool and a non-aqueous restricted proton pool is resolved using a gradient-echo pulsed irradiation sequence that simultaneously encodes the MT and Bloch-Siegert effects for $B_1^+$ inhomogeneity correction. In this current study, we extend BTS by introducing a multi-echo acquisition strategy and a new modeling approach that considers the full spin dynamics among three proton pools: a fast-relaxing aqueous free water pool (e.g., modeling myelin water or cartilage macromolecule-bound water), a slow-relaxing aqueous free water pool (e.g., modeling axonal/extracellular water in brain or bulk water in cartilage) and a non-aqueous restricted proton pool (e.g., modeling brain tissue macromolecules or cartilage extracellular matrix macromolecules). This expanded model, combined with the multi-echo acquisition and relaxed MT parameter assumptions (e.g., allowing fitting for macromolecule pool $T_1$), enables comprehensive characterization of the spin dynamics between these three pools. As a result, the method supports quantification of MT effects while providing a multi-component relaxometry assessment free from potential biases caused by partial longitudinal relaxation, MT effects, or $B_1^+$ inhomogeneity that affect existing methods\cite{Jang2023,Prasloski2012a,Vavasour2000,Sati2013}. Importantly, the multi-echo acquisition does not increase the scan time beyond the original BTS sequence and better leverages its duty cycle. We coin this new approach MC BTS (Multi-Component Bloch-Siegert and magnetization Transfer Simultaneously). In the following sections, we validate an analytically derived signal model for MC BTS using numerical Bloch simulations, introduce a fitting procedure based on this signal equation validated with Monte Carlo simulations and finally demonstrate in vivo feasibility through brain and knee experiments.

\section{THEORY}
\label{sec:theory}
\subsection{Three-pool System}
\begin{figure}
  \centering
  \includegraphics{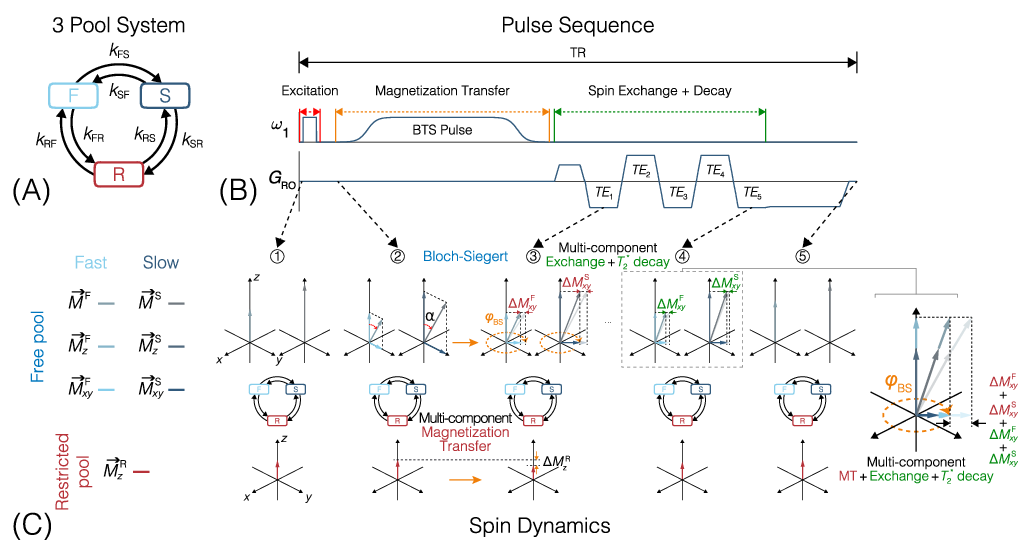}
  \caption{(a) Three-pool system composed of two “free” liquid proton pools, one fast-relaxing (F) and the other slow-relaxing (S), and a “restricted” macromolecule proton (R), where the magnetization exchange between these three pools is characterized by pseudo-first-order rates. (b) MC BTS acquisition scheme is composed of a magnetization transfer module (orange) and a spin exchange + decay module (green). When the BTS criteria are satisfied, in the magnetization transfer module, (c) transverse magnetization of the “free” liquid pools acquires a phase proportional to the peak magnitude of the BTS pulse. Concurrently, the “restricted” pool is partially saturated, which is reflected as a decrease in the observable transverse magnetization of the multicomponent “free” liquid pools ($\Delta M_\text{xy}^\text{F}$, $\Delta M_\text{xy}^\text{S}$). In the spin exchange + decay module, there is a constant exchange of spins between pools F and S while both pools undergo $T_2^*$ decay.}
  \label{fig1}
\end{figure}

Let us consider a three-pool spin system composed of two “free” liquid proton pools, one fast-relaxing and the other slow-relaxing, denoted by F and S, and a “restricted” macromolecule proton pool denoted by R (Figure \ref{fig1}A). The magnetization exchange between these three pools can be characterized by pseudo-first-order rates ($k_\text{FR}$, $k_\text{RF}$, $k_\text{SR}$, $k_\text{RS}$, $k_\text{FS}$ and $k_\text{SF}$) where the first letter of the subscript denotes the origin and the second letter the destination of exchange (e.g., $k_\text{FR}$ is the magnetization exchange rate from F to R and $k_\text{RF}$ is vice versa). In equilibrium, $k_\text{FR} = k_1M_0^\text{R}$, $k_\text{RF} = k_1M_0^\text{F}$, $k_\text{SR} = k_2M_0^\text{R}$, $k_\text{RS} = k_2M_0^\text{S}$, $k_\text{FS} = k_3M_0^\text{S}$ and $k_\text{SF} = k_3M_0^\text{F}$, where $k_1$, $k_2$ and $k_3$ are the fundamental exchange rate constants between pools F and R, S and R, and F and S, respectively, and $M_0^\text{F}$, $M_0^\text{S}$ and $M_0^\text{R}$ are the corresponding equilibrium magnetizations of the fast-relaxing, slow-relaxing and restricted pools.\newline

The dynamics of the spin magnetization exchange between the three pools are governed by the Bloch-McConnell equations\cite{McConnell1958}. Its general form is given by

\begin{equation} \label{eq:1}
\frac{\partial M_\text{x}^\text{F} (t)}{\partial t}=-R_2^\text{F} M_\text{x}^\text{F}(t)-k_\text{FS} M_\text{x}^\text{F} (t)+k_\text{SF} M_\text{x}^\text{S}(t)+\Delta\omega_\text{F} M_\text{y}^\text{F}(t)-\Im{\{\omega_1(t)\}} M_\text{z}^\text{F}(t)
\end{equation}
\begin{equation} \label{eq:2}
\frac{\partial M_\text{y}^\text{F} (t)}{\partial t}=-R_2^\text{F} M_\text{y}^\text{F}(t)-k_\text{FS} M_\text{y}^\text{F} (t)+k_\text{SF} M_\text{y}^\text{S}(t)-\Delta\omega_\text{F} M_\text{x}^\text{F}(t)+\Re{\{\omega_1(t)\}} M_\text{z}^\text{F}(t)
\end{equation}
\begin{equation} \label{eq:3}
\frac{\partial M_\text{x}^\text{S} (t)}{\partial t}=-R_2^\text{S} M_\text{x}^\text{S}(t)-k_\text{SF} M_\text{x}^\text{S} (t)+k_\text{FS} M_\text{x}^\text{F}(t)+\Delta\omega_\text{S} M_\text{y}^\text{S}(t)-\Im{\{\omega_1(t)\}} M_\text{z}^\text{S}(t)
\end{equation}
\begin{equation} \label{eq:4}
\frac{\partial M_\text{y}^\text{S} (t)}{\partial t}=-R_2^\text{S} M_\text{y}^\text{S}(t)-k_\text{SF} M_\text{y}^\text{S} (t)+k_\text{FS} M_\text{y}^\text{F}(t)-\Delta\omega_\text{S} M_\text{x}^\text{S}(t)+\Re{\{\omega_1(t)\}} M_\text{z}^\text{S}(t)
\end{equation}
\begin{equation} \label{eq:5}
\frac{\partial M_\text{z}^\text{F} (t)}{\partial t}=-R_1^\text{F} (M_\text{z}^\text{F}(t)- M_0^\text{F})-(k_\text{FR} + k_\text{FS})M_\text{z}^\text{F} (t)+k_\text{SF} M_\text{z}^\text{S}(t)+k_\text{RF} M_\text{z}^\text{R}(t) - \Re{\{\omega_1(t)\}} M_\text{y}^\text{F}(t) + \Im{\{\omega_1(t)\}} M_\text{x}^\text{F}(t)
\end{equation}
\begin{equation} \label{eq:6}
\frac{\partial M_\text{z}^\text{S} (t)}{\partial t}=-R_1^\text{S} (M_\text{z}^\text{S}(t)- M_0^\text{S})-(k_\text{SR} + k_\text{SF})M_\text{z}^\text{S}(t)+k_\text{FS} M_\text{z}^\text{F}(t)+k_\text{RS} M_\text{z}^\text{R}(t) - \Re{\{\omega_1(t)\}} M_\text{y}^\text{S}(t) + \Im{\{\omega_1(t)\}} M_\text{x}^\text{S}(t)
\end{equation}
\begin{equation} \label{eq:7}
\frac{\partial M_\text{z}^\text{R} (t)}{\partial t}=-R_1^\text{R} (M_\text{z}^\text{R}(t)- M_0^\text{R})-(k_\text{RF} + k_\text{RS})M_\text{z}^\text{R}(t)+k_\text{FR} M_\text{z}^\text{F}(t)+k_\text{SR} M_\text{z}^\text{S}(t) - \langle W(\Delta) \rangle M_\text{z}^\text{R}(t)
\end{equation}

where $R_1^\text{F[S,R]} = 1/T_1^\text{F[S,R]}$ are the corresponding spin-lattice relaxation rates of the fast-relaxing, slow-relaxing and restricted macromolecule pools, respectively, and $R_2^\text{F[S]} = 1/T_2^\text{F[S]}$ are the spin-spin relaxation rates of the fast- and slow-relaxing molecule free pool. $\Delta\omega_\text{F[S]}$ entails any common off-resonance contributions such as static field ($B_0$) inhomogeneity and/or variations in the applied RF frequency, in addition to pool-specific influences originating from its microenvironment. $\omega_1(t)$ is the time-dependent RF amplitude and $\langle W(\Delta)\rangle$ is the average saturation rate of the macromolecule pool due to irradiation applied at frequency offset $\Delta$, which is proportional to both the macromolecule absorption line shape and the square of applied RF amplitude\cite{Graham1997}, given by

\begin{equation} \label{eq:8}
\langle W(\Delta)\rangle=\pi\frac{1}{\tau_\text{RF}} \int_{0}^{\tau_\text{RF}}\omega_1^2(t)dtG(\Delta)
\end{equation}

where $\tau_\text{RF}$ is the irradiating pulse width, $\omega_1(t)$ is the time-dependent pulse amplitude and $G(\Delta)$ is the frequency-dependent absorption line shape at an offset $\Delta$ with respect to the resonance frequency. Here, we employ the Super-Lorentzian absorption line shape which adequately characterizes saturation of macromolecules in tissues such as white matter (WM), grey matter (GM) and cartilage\cite{Liu2016a,Morrison1995}

\begin{equation} \label{eq:9}
G_\text{SL}(\Delta)=\int_{0}^{1}\sqrt{\frac{2}{\pi}}\frac{T_2^\text{R}}{|3u^2-1|}e^{-2\big[\frac{2\pi\Delta T_2^\text{R}}{3u^2-1}\big]^2}du
\end{equation}

where the on-resonance singularity was estimated through extrapolation from $\Delta$ = 1kHz to the asymptotic limit $\Delta \rightarrow$ 0, giving 1.4x10$^{-5}$ [s$^{-1}$]\cite{Bieri2006}.

\subsubsection{MC BTS Signal Model}
The MC BTS acquisition sequence, which is based on an RF-spoiled gradient-echo scheme, is presented in Figure \ref{fig1}B. In contrast to selectively saturating the restricted pool prior to excitation as a preparation module, as is commonly practiced\cite{Sled2001,Henkelman1993}, off-resonance irradiation instead is realized between excitation and acquisition. If the BTS (Bloch-Siegert and magnetization Transfer Simultaneously) criteria\cite{Jang2023} are satisfied, specifically, when the offset frequency of the applied pulse is large relative to its peak amplitude and the center of its saturation profile is sufficiently shifted away from the center of the free pool absorption lineshape such that direct saturation can be neglected, this allows partial saturation of the broad absorption lineshape of the macromolecular pool with minimal saturation of the liquid pools, while simultaneously satisfying the conditions necessary to generate a $B_1^+$-dependent Bloch-Siegert phase shift\cite{Sacolick2010}. Accordingly, in steady-state, this partial saturation of the macromolecule gets “transferred” as a decrease in the observable transverse magnetization of the two free pools, which in tandem obtains a phase proportional to transmit $B_1^+$ (Figure \ref{fig1}C).\newline

An analytical signal model can be derived using Eqs. [\ref{eq:1}] – [\ref{eq:7}], which can be rearranged into the following matrix form:

\begin{equation} \label{eq:10}
\frac{\partial \vec{M}(t)}{\partial t}=\textbf{A}\vec{M}(t)+\vec{M}_\text{eq}
\end{equation}

\begin{equation} \label{eq:11}
\textbf{A}=\begin{bmatrix}
\textbf{A}_\textbf{xy} & 0 \\
0 & \textbf{A}_\textbf{z}
\end{bmatrix}
\end{equation}

\begin{equation} \label{eq:12}
\textbf{A}_\textbf{xy}=\begin{bmatrix}
-R_2^\text{F} - k_\text{FS} & \Delta\omega_\text{F} & k_\text{SF} & 0 \\
-\Delta\omega_\text{F} & -R_2^\text{F} - k_\text{FS} & 0 & k_\text{SF} \\
k_\text{FS} & 0 & -R_2^\text{S} - k_\text{SF} & \Delta\omega_\text{S} \\
0 & k_\text{FS} & -\Delta\omega_\text{S} & -R_2^\text{S} - k_\text{SF}
\end{bmatrix}
\end{equation}

\begin{equation} \label{eq:13}
\textbf{A}_\textbf{xy}=\begin{bmatrix}
-R_1^\text{F} - k_\text{FR} - k_\text{FS} & k_\text{SF} & k_\text{RF} \\
k_\text{FS} & -R_1^\text{S} - k_\text{SR} - k_\text{SF} & k_\text{RS}\\
k_\text{FR} & k_\text{SR} & -R_1^\text{R} - k_\text{RF} - k_\text{RS} - \langle W(\Delta)\rangle
\end{bmatrix}
\end{equation}

In the above, we assume $\Delta = \Delta_\text{off}$ satisfies the BTS condition and excitation occurs instantaneously at the beginning of the repetition interval, allowing omission of the $\omega_1(t)$ terms. $\vec{M}(t)=\begin{bmatrix} M_\text{x}^\text{F}(t) & M_\text{y}^\text{F}(t) & M_\text{x}^\text{S}(t) & M_\text{y}^\text{S}(t) & M_\text{z}^\text{F}(t) & M_\text{z}^\text{S}(t) & M_\text{z}^\text{R}(t)\end{bmatrix}^\text{T}$ is the time dependent vector that comprises the transverse and longitudinal components of each pool, $\vec{M}_\text{eq}=M_0 \begin{bmatrix} 0 & 0 & 0 & 0 & R_1^\text{F}f_\text{F} & R_1^\text{S}f_\text{S} & R_1^\text{R}f_\text{R} \end{bmatrix}^\text{T}$ is the equilibrium vector where $f_\text{F}$, $f_\text{S}$ and $f_\text{R}$ correspond to fast-relaxing, slow-relaxing and macromolecular proton fraction with respect to the total magnetization $M_0$ and add up to 1 ($M_0^\text{F} = f_\text{F}M_0$, $M_0^\text{S} = f_\text{S}M_0$, $M_0^\text{R} = f_\text{R}M_0$).

For RF-spoiled gradient-echo, assuming perfect spoiling of the transverse magnetization at the end of each TR, to derive the steady-state magnetization, we only consider the longitudinal magnetization components

\begin{equation} \label{eq:14}
\frac{\partial \vec{M}_\text{z}(t)}{\partial t}=\textbf{A}_\textbf{z}\vec{M}_\text{z}(t)+\vec{M}_\text{z,eq}
\end{equation}

$\vec{M}_\text{z}(t)=\begin{bmatrix} M_\text{z}^\text{F}(t) & M_\text{z}^\text{S}(t) & M_\text{z}^\text{R}(t)\end{bmatrix}^\text{T}$ is the time dependent longitudinal vector of each pool, $\vec{M}_\text{z,eq}=M_0 \begin{bmatrix} R_1^\text{F} f_\text{F} & R_1^\text{S} f_\text{S} & R_1^\text{R} f_\text{R}\end{bmatrix}^\text{T}$ is the longitudinal equilibrium vector. The general solution for Eq. [14] is given by

\begin{equation} \label{eq:15}
\vec{M}_\text{z}(t)=e^{\textbf{A}_\textbf{z}t}(\vec{M}_\text{z}(t=0)+\textbf{A}_\textbf{z}^{-1}\vec{M}_\text{z,eq}) - \textbf{A}_\textbf{z}^{-1}\vec{M}_\text{z,eq}
\end{equation}

For MC BTS, in the (n + 1)$^\text{th}$ repetition, the magnetization after off-resonance saturation becomes

\begin{equation} \label{eq:16}
\vec{M}_\text{z}(nTR+\tau_\text{BTS})=e^{\textbf{A}_\textbf{z}\tau_\text{BTS}}(\vec{M}_\text{z}^+(nTR)+\textbf{A}_\textbf{z}^{-1}\vec{M}_\text{z,eq}) - \textbf{A}_\textbf{z}^{-1}\vec{M}_\text{z,eq}
\end{equation}

where $\vec{M}_z^+(nTR)$ is the magnetization state following the (n + 1)$^\text{th}$ excitation using an excitation pulse length $\tau_\text{exc}$ and $\tau_\text{BTS}$ is the BTS inducing pulse length. At the end of the (n + 1)$^\text{th}$ repetition, the magnetization state $\vec{M}_\text{z}^-((n+1)TR)$ is given by

\begin{equation} \label{eq:17}
\vec{M}_\text{z}^-((n+1)TR)=e^{\textbf{A}_\textbf{z0}(TR-\tau_\text{BTS})}(\vec{M}_\text{z}(nTR+\tau_\text{BTS})+\textbf{A}_\textbf{z0}^{-1}\vec{M}_\text{z,eq}) - \textbf{A}_\textbf{z0}^{-1}\vec{M}_\text{z,eq}
\end{equation}

where $\textbf{A}_\textbf{z0}$ is $\textbf{A}_\textbf{z}$ with $\langle W(\Delta)\rangle$ set to 0, reflecting only relaxation and spin exchange effects. Substituting Eq. [16] into Eq. [17] along with $\vec{M}_\text{z}^+(nTR)=\textbf{R}_\textbf{exc}(\alpha)\vec{M}_\text{z}^-(nTR)$, where

\begin{equation} \label{eq:18}
\textbf{R}_\textbf{exc}(\alpha)=\begin{bmatrix}
\cos(\alpha) & 0 & 0 \\
0 & \cos(\alpha) & 0 \\
0 & 0 & e^{-\langle W(0) \rangle\tau_\text{exc}} \\
\end{bmatrix}
\end{equation}

is the excitation matrix assuming RF is applied along the $x$-axis with on-resonance saturation included, one can solve for the steady-state condition $\vec{M}_\text{z}^-((n+1)TR)=\vec{M}_\text{z}^-(nTR)=\vec{M} _\text{z,ss}=\begin{bmatrix} M_\text{z,ss}^\text{F} & M_\text{z,ss}^\text{S} & M_\text{z,ss}^\text{R}\end{bmatrix}^\text{T}$ accordingly

\begin{equation} \label{eq:19}
\vec{M}_\text{z,ss}=[\textbf{I}_3-e^{\textbf{A}_\textbf{z0}(TR-\tau_\text{BTS})}e^{\textbf{A}_\textbf{z}\tau_\text{BTS}}\textbf{R}_\textbf{exc}(\alpha)]^{-1}[e^{\textbf{A}_\textbf{z0}(TR-\tau_\text{BTS})}(e^{\textbf{A}_\textbf{z}\tau_\text{BTS}}-\textbf{I}_3)\textbf{A}_\textbf{z}^{-1}+(e^{\textbf{A}_\textbf{z0}(TR-\tau_\text{BTS})}-\textbf{I}_3)\textbf{A}_\textbf{z0}^{-1}]\vec{M}_\text{z,eq}
\end{equation}

where $\textbf{I}_3$ is the 3x3 identity matrix.\newline

Subsequently, the observable complex signal at time $TE$ after excitation can be extracted from the steady-state magnetization

\begin{equation} \label{eq:20}
\begin{bmatrix} M_\text{xy,ss}^\text{F}(TE) \\ M_\text{xy,ss}^\text{S}(TE)\end{bmatrix}=
e^{\textbf{A}_\textbf{xy}TE}
\begin{bmatrix}
M_\text{z,ss}^\text{F}\sin(\alpha)e^{i\phi_\text{BTS}} \\
M_\text{z,ss}^\text{S}\sin(\alpha)e^{i\phi_\text{BTS}}
\end{bmatrix}
\end{equation}

$\phi_\text{BTS}=B_\text{1,max}^2 \int_0^{\tau_\text{BTS}}\frac{(\gamma B_\text{1,normalized}(t))^2}{2(2\pi\Delta_\text{off})}dt=B_\text{1,max}^2 K_\text{BS}$, where $B_\text{1,max}$ and $B_\text{1,normalized}(t)$ are the peak amplitude and amplitude modulation function normalized to 1, respectively, of the BTS pulse applied at offset frequency $\Delta_\text{off}$, is the Bloch-Siegert phenomena-induced phase. Note that Eq. [20] accounts for both decay and spin exchange that takes place after excitation for a time period $TE$. For RF-spoiled gradient-echo, the spin-spin relaxation terms $R_2^\text{F}$ and $R_2^\text{S}$ in $\textbf{A}_\textbf{xy}$ are the effective spin-spin relaxation terms $R_2^\text{F*}$ and $R_2^\text{S*}$.

For the case when the BTS pulse is not applied (referred to as baseline [BL]), the steady-state condition can be solved using Eq [15] replacing $\textbf{A}_\textbf{z}$ with $\textbf{A}_\textbf{z0}$ to reflect the absence of off-resonance BTS application

\begin{equation} \label{eq:21}
\vec{M}_\text{z,ss}^\text{BL}=[\textbf{I}_3-e^{\textbf{A}_\textbf{z0}TR}\textbf{R}_\textbf{exc}(\alpha)]^{-1}[e^{\textbf{A}_\textbf{z0}TR}-\textbf{I}_3)]\textbf{A}_\textbf{z0}^{-1}\vec{M}_\text{z,eq}
\end{equation}

\begin{equation} \label{eq:22}
\begin{bmatrix} M_\text{xy,ss}^\text{F,BL}(TE) \\ M_\text{xy,ss}^\text{S,BL}(TE)\end{bmatrix}=
e^{\textbf{A}_\textbf{xy}TE}
\begin{bmatrix}
M_\text{z,ss}^\text{F,BL}\sin(\alpha) \\
M_\text{z,ss}^\text{S,BL}\sin(\alpha)
\end{bmatrix}
\end{equation}

where $\vec{M} _\text{z,ss}^\text{BL}=\begin{bmatrix} M_\text{z,ss}^\text{F,BL} & M_\text{z,ss}^\text{S,BL} & M_\text{z,ss}^\text{R,BL}\end{bmatrix}^\text{T}$.

\section{METHODS}
\subsection{MC BTS Parameter Estimation}
\begin{figure}
  \centering
  \includegraphics{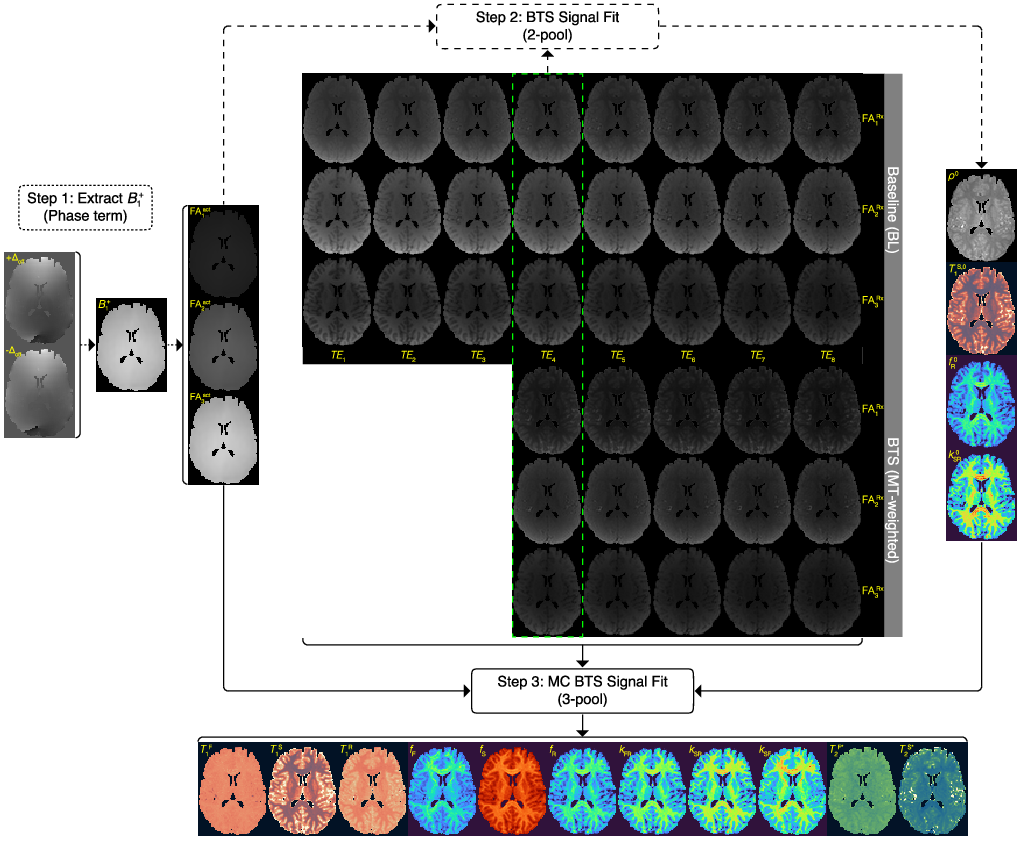}
  \caption{Multiparametric estimation pipeline of MC BTS consists of 3 steps utilizing multi-echo acquisitions obtained with (BTS) and without (BL) off-resonance BTS pulse applied at multiple excitation angles. Step 1: Utilizing an additional BTS acquisition acquired at an excitation angle closest to the Ernst angle, any $\Delta B_0$ inhomogeneity and chemical shift originating phase is removed to extract Bloch-Siegert induced phase shift to derive a $B_1^+$ map and corresponding actual flip angle maps (FA$_\text{n}^\text{act}$). Step 2: The spatially varying actual FA maps, BTS and BL magnitude images acquired at the shortest identical echo time (TE$_4$ in this example) are used to fit the two-pool BTS signal model, generating $\rho$, $T_1^\text{S}$, $f_\text{R}$ and $k_\text{SR}$ maps. Step 3: Utilizing the estimated two-pool parameters as a prior, the multi-echo BTS and BL data obtained from multiple flip angles together with the actual FA maps are combined to estimate the three-pool parameters ($T_1^\text{F}$, $T_1^\text{S}$, $T_1^\text{R}$, $f_\text{F}$, $f_\text{S}$, $f_\text{R}$, $k_\text{FR}$, $k_\text{SR}$, $k_\text{SF}$, $T_2^\text{F*}$, $T_2^\text{S*}$) using the MC BTS signal model (Eqs. [\ref{eq:20}] and [\ref{eq:22}]).}
  \label{fig2}
\end{figure}

Two sets of data, one without (i.e., BL) and one with the BTS pulse applied, each at multiple excitation flip angles and multiple echo times, are collected as MC BTS data. The multiparametric estimation pipeline of MC BTS consists of 3 steps. Referring to Figure \ref{fig2}, in the first step, to remove any $B_0$ inhomogeneity and chemical shift dependence on the Bloch-Siegert shift, an additional BTS acquisition with the BTS pulse offset frequency applied symmetrically about the carrier frequency at a flip angle closest to the Ernst angle is performed. From this, a $B_1^+$ map is extracted from the phase image, which are subsequently utilized to generate actual FA maps that reflect $B_1$ inhomogeneity. In step two, the spatially varying actual FA maps are combined with both the BL and BTS magnitude images obtained from an identical echo time (i.e., TE$_4$) and used to fit the magnitude term of the original two-pool BTS signal equation36 pixel by pixel, thus generating $T_1^\text{S,0}$, $f_\text{R}^0$ and $k_\text{F[R]}^0$ maps, assuming $T_1^\text{R}$ = 1 s and $T_2^\text{R}$ = 11 $\mu$s\cite{Yarnykh2004,Sled2001,Gloor2008,Henkelman1993}. In step three, the estimated two-pool parameters are utilized as initial values to fit the BL and BTS magnitude images obtained from multiple echoes together with the spatially varying actual FA maps to the three-pool MC BTS signal equation (Eqs. [\ref{eq:19}] – [\ref{eq:22}]) pixel by pixel, generating $T_1$, $f$ and $k$ maps of the fast-relaxing, slow-relaxing and restricted pools, along with $T_2^*$ of the fast- and slow-relaxing pools, assuming $T_2^\text{R}$ = 11 $\mu$s\cite{Yarnykh2004,Sled2001,Gloor2008,Henkelman1993}.\newline

\subsection{Simulations}
The MC BTS signal equations both with (Eqs. [\ref{eq:19}] – [\ref{eq:20}]) and without (Eqs. [\ref{eq:21}] – [\ref{eq:22}]) BTS applied was compared and verified with an in-house numerical Bloch simulator programmed in C language and interfaced with MATLAB (MathWorks, Inc., Natick, MA) which incrementally solved Eqs. [\ref{eq:1}] – [\ref{eq:7}] in 2 $\mu$s time increments with the actual RF waveform discretized\cite{Liu2017a}. For all three pools, tissue parameters that were previously reported at 3T within a series of studies\cite{Yarnykh2004,Jang2023,West2019,Wang2020,Lee2021} were used: $f_\text{F}$ = 21.7\%\cite{West2019}, $f_\text{S}$ = 65.1\%, $f_\text{R}$ = 13.2\%\cite{Jang2023}, $k_\text{FR}$ = 1.5 s$^{-1}$\cite{Jang2023}, $k_\text{SR}$ = 3.1 s$^{-1}$\cite{Jang2023}, $k_\text{SF}$ = 3.33 s$^{-1}$\cite{West2019}, $T_1^\text{F}$ = 350 ms\cite{West2019}, $T_1^\text{S}$ = 800 ms\cite{Lee2021}, $T_1^\text{R}$ = 257 ms\cite{Wang2020}, $T_2^\text{F*}$ = 10 ms\cite{Lee2021}, $T_2^\text{S*}$ = 40 ms\cite{Lee2021}, $T_2^\text{R}$ = 11 $\mu$s\cite{Yarnykh2004}. Common sequence parameters used in both BL and MC BTS simulations were TR = 45 ms and $\tau_\text{exc}$ = 0.5 ms. The BL simulations sampled 8 data points at TE = 3.25 $\sim$ 22.5 ms in 2.75 ms increments whereas the MC BTS simulations sampled 5 data points at TE = 11.5 $\sim$ 22.5 ms in 2.75 ms increments. For MC BTS simulations, a fermi pulse of $\tau_\text{BTS}$ = 8 ms, $B_\text{1,max}$ = 7.3 $\mu$T ($B_\text{1,rms}$ = 5.93 $\mu$T) and $\Delta_\text{off}$ = 4000 Hz with an accompanying Super-Lorentzian absorption line shape (Eq. [9]) was used, following a similar setup of a previous study\cite{Jang2023}. The excitation flip angle was swept from 1$^\circ$ $\sim$ 90$^\circ$ in 1$^\circ$ increments. All simulation data were sampled from the steady-state after 250 repetitions.\newline

To analyze the accuracy and stability of the MC BTS parameter estimation pipeline, 3 sets of Monte Carlo simulations (n = 50,000 observations per set), each with varying levels of signal-to-noise ratio (SNR), were carried out. Simulation of the three-pool model was conducted using identical tissue and sequence parameters above except for the sampled flip angles which were 7$^\circ$, 18$^\circ$ and 40$^\circ$. This was followed by the addition of white Gaussian noise of standard deviation 0.0035, 0.0018 and 0.0009 in units of thermal equilibrium magnetization, resulting in SNRs of 25, 50 and 100 with respect to the mean signal value. For each set, the generated BL and BTS data were used to carry out multiparametric estimation using the MC BTS signal model following the method previously described.

\subsection{Experiments}
\subsubsection{In Vivo Brain}
In vivo study of the human brain was carried out on five healthy volunteers (two males, three females, ages 36-41 years) to test the feasibility of the MC BTS method under a protocol approved by our institution’s institutional review board. Experiments were carried out on a 3T clinical scanner (Prisma Fit, software version XA30, Siemens Healthineers, Erlangen, Germany) equipped with a 64-channel head receiver. Common 3D sequence parameters used in both BL and MC BTS measurements were: matrix size = 208x106x80, yielding 1.1x2.2x2.2 mm resolution in the sagittal view and TR = 42 ms. 10 dummy scans and a reverse centric $k$-space acquisition scheme were carried out to ensure contrast was acquired in steady-state. 169$^\circ$ linear RF phase increments between subsequent repetition cycles and strong gradient spoilers (0$^\text{th}$ order moment = 510 mT$\cdot$ms/m) at the end of each repetition cycle were applied to minimize the impact of imperfect spoiling on $T_1$ measurement\cite{Preibisch2009,Yarnykh2010} and Bloch-Siegert shift\cite{Corbin2019}. For MC BTS, the same fermi pulse, and for BL and MC BTS, the same prescription flip angles (7$^\circ$, 18$^\circ$, 40$^\circ$) and echo times used in the simulations were measured based on a monopolar scheme. For both BL and MC BTS acquisitions, parallel imaging was applied along the phase encoding direction using an acceleration factor of 2 with 24 calibration lines, resulting in a total scan time of 27 min.

Scans for the $T_2$-based method, which do not model MT effects, were additionally carried out for comparison. For the $T_2$-based method, a 3D multi-echo spin-echo using GRASE readout was used to acquire a matrix size = 128x112x70 yielding 1.8x1.8x2.2 mm resolution in the axial view for 32 echoes (TE = 10 – 320 ms with 10 ms echo spacing\cite{Prasloski2012}) using TR = 1 s. The acquired data was processed to extract the relative portion of the fast-relaxing and slow-relaxing pool using regularized non-negative least squares\cite{Canales-Rodrguez2021}, where generalized cross validation was applied to estimate the regularization parameter.

\subsubsection{In Vivo Knee}
In vivo study of the knee was carried out on five healthy volunteers (two males, three females, ages 36-41 years) to test the applicability of our method to other tissue types. Experiments were performed on the same system used for the brain experiments using a 15-channel transmit/receive knee coil. Common 3D sequence parameters used in both BL and BTS acquisitions were: matrix size = 240x200x36 yielding 0.7x0.7x3 mm resolution along the sagittal orientation and TR = 37 ms. 10 dummy scans and a reverse centric $k$-space acquisition scheme were carried out to ensure contrast was acquired in steady-state. 169$^\circ$ linear RF phase increments between subsequent repetition cycles and strong gradient spoilers (0$^\text{th}$ order moment = 310 mT$\cdot$ms/m) at the end of each repetition cycle were applied to minimize the impact of imperfect spoiling\cite{Preibisch2009,Yarnykh2010}. For MC BTS, the same fermi pulse, and for BL and MC BTS, the same echo times used in the simulations were measured based on a monopolar scheme using prescription flip angles 10$^\circ$, 20$^\circ$, 40$^\circ$, resulting in a total scan time of 31 min.

\begin{table}[ht]
\caption{}
\centering
\resizebox{\textwidth}{!}{\begin{tabular}{lccccccccccccc} 
\hhline{==============}
\multirow{2}{*}{Protocol}                                              & \multirow{2}{*}{Orientation} & \multirow{2}{*}{FOV [mm]}    & \multirow{2}{*}{\begin{tabular}[c]{@{}c@{}}Resolution \\{[}mm]\end{tabular}} & \multirow{2}{*}{TR [ms]} & \multirow{2}{*}{TE1 [ms]} & \multirow{2}{*}{\begin{tabular}[c]{@{}c@{}}echo spacing \\{[}ms]\end{tabular}} & \multirow{2}{*}{\# of echoes} & \multirow{2}{*}{FA [deg]}      & \multicolumn{3}{c}{BTS Pulse}                                         & \multirow{2}{*}{Parallel Imaging}                                                 & \multirow{2}{*}{Scan Time [min]}  \\ 
\cline{10-12}
                                                                       &                              &                              &                                                                              &                          &                           &                                                                                &                               &                                & type                   & length [ms]        & $B_\text{1,max}$                  &                                                                                   &                                   \\ 
\hline\hline
\multirow{2}{*}{\begin{tabular}[c]{@{}l@{}}MC BTS\\Knee\end{tabular}}  & \multirow{2}{*}{SAG}         & \multirow{2}{*}{168x140x108} & \multirow{2}{*}{0.7x0.7x3.0}                                                 & \multirow{2}{*}{37}      & MC BTS: 12                & \multirow{2}{*}{3.2}                                                           & MC BTS: 5                     & \multirow{2}{*}{10°, 20°, 40°} & \multirow{2}{*}{Fermi} & \multirow{2}{*}{8} & \multirow{2}{*}{7.3 $\mu$T} & \multirow{2}{*}{None}                                                             & \multirow{2}{*}{31}               \\
                                                                       &                              &                              &                                                                              &                          & BL: 2.9                   &                                                                                & BL: 8                         &                                &                        &                    &                         &                                                                                   &                                   \\
\multirow{2}{*}{\begin{tabular}[c]{@{}l@{}}MC BTS\\Brain\end{tabular}} & \multirow{2}{*}{SAG}         & \multirow{2}{*}{228x233x176} & \multirow{2}{*}{1.1x2.2x2.2}                                                 & \multirow{2}{*}{42}      & MC BTS: 11.5              & \multirow{2}{*}{2.75}                                                          & MC BTS: 5                     & \multirow{2}{*}{7°, 18°, 40°}  & \multirow{2}{*}{Fermi} & \multirow{2}{*}{8} & \multirow{2}{*}{7.3 $\mu$T} & \multirow{2}{*}{\begin{tabular}[c]{@{}c@{}}Acceleration\\Factor = 2\end{tabular}} & \multirow{2}{*}{27}               \\
                                                                       &                              &                              &                                                                              &                          & BL: 3.25                  &                                                                                & BL: 8                         &                                &                        &                    &                         &                                                                                   &                                   \\
\begin{tabular}[c]{@{}l@{}}ME-SE \\GRASE\end{tabular}                  & AXL                          & 230x202x154                  & 1.8x1.8x2.2                                                                  & 1000                     & 10                        & 10                                                                             & 32                            & N/A                            & N/A                    & N/A                & N/A                     & \begin{tabular}[c]{@{}c@{}}Acceleration \\Factor = 6\end{tabular}                 & 9.5                              
\end{tabular}}
  \label{table1}
\end{table}

\subsubsection{Regional analysis}
Regional analysis of the brain was performed using segmentation software\cite{Billot2023} to identify WM and GM regions. Representative regions of interest (ROI) in both WM and GM were chosen to validate our method. For the knee, manual segmentation of the patella, trochlea and lateral tibia plateau was carried out and chosen as ROIs for cartilage. In these regions, the mean and SD of the MT parameters $T_1$, $f$ and $k$ for the three pools (fast-relaxing, slow-relaxing and restricted), along with $T_2^*$ for the fast-relaxing and slow-relaxing pools were evaluated.

\section{RESULTS}
\subsection{Simulations}
\begin{figure}
  \centering
  \includegraphics{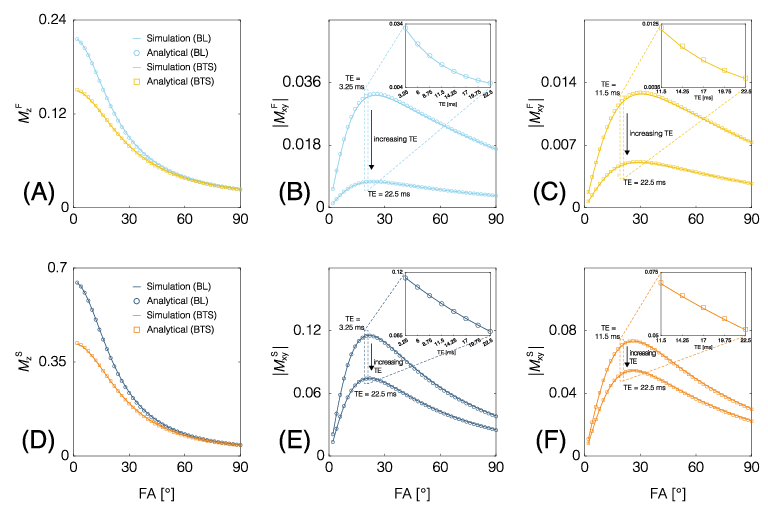}
  \caption{Simulation results overlaid with MC BTS signal model results for varying excitation flip angles are shown for the fast-relaxing pool (top row) and slow-relaxing pool (bottom row). For the fast-relaxing pool, simulation results (solid lines) of (a) longitudinal and transverse magnetization for (b) BL (light blue) and (c) BTS (yellow) show excellent agreement with the MC BTS signal model (circle and square markers). Further simulating across varying echo times shown by the dashed box in (b) and (c), where echo time increases going from top to down, excellent agreement between simulation and signal model is also observed, as can be seen in the figure inlets. The same applies for the slow-relaxing pool (d) – (f), where BL is color coded in navy and BTS in orange.}
  \label{fig3}
\end{figure}

Simulations carried out with $\Delta_\text{off}$ satisfying the BTS condition and BL, each overlayed with its corresponding analytical equation (Eqs. [\ref{eq:19}] – [\ref{eq:22}]), are presented in Figure \ref{fig3}. Within experimentally feasible flip angles (1$^\circ$ $\sim$ 40$^\circ$), the maximum deviation between the simulation data and the analytical equation of the longitudinal magnetization for BTS (Figures 3a and 3d) is 1.1\% and 1.0\% for the fast-relaxing and slow-relaxing pool, respectively. For BL, the maximum deviation is 0.5\% and 0.4\% for the fast-relaxing and slow-relaxing pool, respectively. At echo times varying from (11.5 $\sim$ 22.5 ms for BTS and 3.25 $\sim$ 22.5 ms for BL), the maximum deviation between the simulation data and the analytical equation of the transverse magnetization is 3.1\% and 1.6\% for the fast-relaxing and slow-relaxing pool using BTS (Figures 3c and 3f), and 2.5\% and 0.7\% for BL (Figures 3b and 3e). Overall, the analytical equation and simulation show excellent agreement.

\begin{figure}
  \centering
  \includegraphics[width=1.0\textwidth]{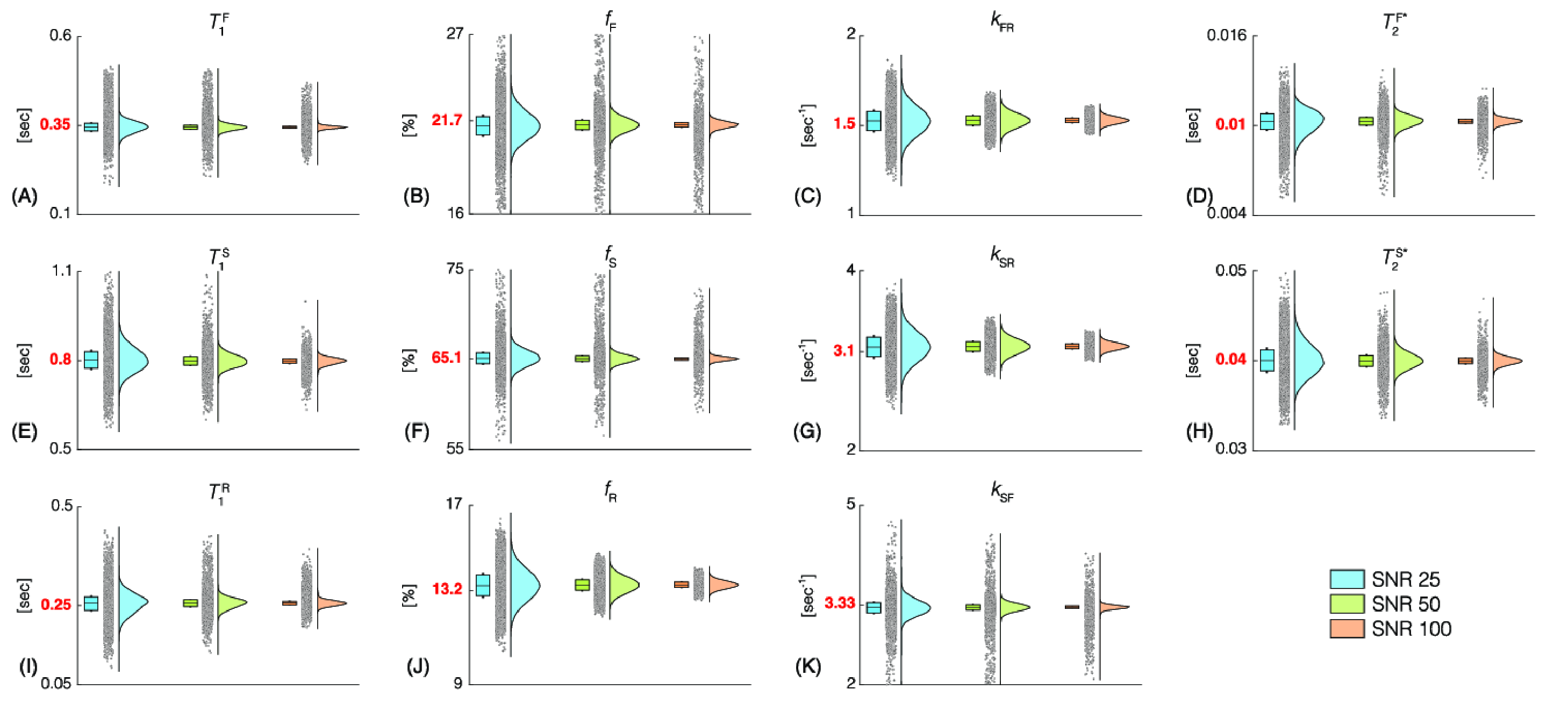}
  \caption{MC BTS parameter estimation results from three sets of Monte Carlo simulations to validate the MC BTS parameter estimation pipeline, each generated using SNR of 25 (sky blue), 50 (light green) and 100 (light orange). $T_1$ (column 1), relative fraction (column 2) and exchange rate (column 3) of the fast-relaxing pool (top row), slow-relaxing pool (middle row) and restricted pool (bottom row), along with $T_2^*$ of the fast and slow-relaxing pools (column 4) agree well with their respective ground truth for all three sets, demonstrating our fitting procedure’s robustness over a wide range of SNR settings. Increasing SNR decreases its estimated variance due to decreased noise.}
  \label{fig4}
\end{figure}

Monte Carlo simulation results are presented in Figure \ref{fig4}. For all parameters, increasing SNR decreases its estimated variance due to decreased noise, as expected. Nonetheless, for all three SNR settings, MC BTS produced parameter estimates for the three pools whose means are in close agreement with each other and the ground truth. This demonstrates the robustness of the MC BTS estimation pipeline, performing well over a wide range of SNR settings.

\subsection{Experiments}
\subsubsection{In Vivo Brain}
\begin{figure}
  \centering
  \includegraphics{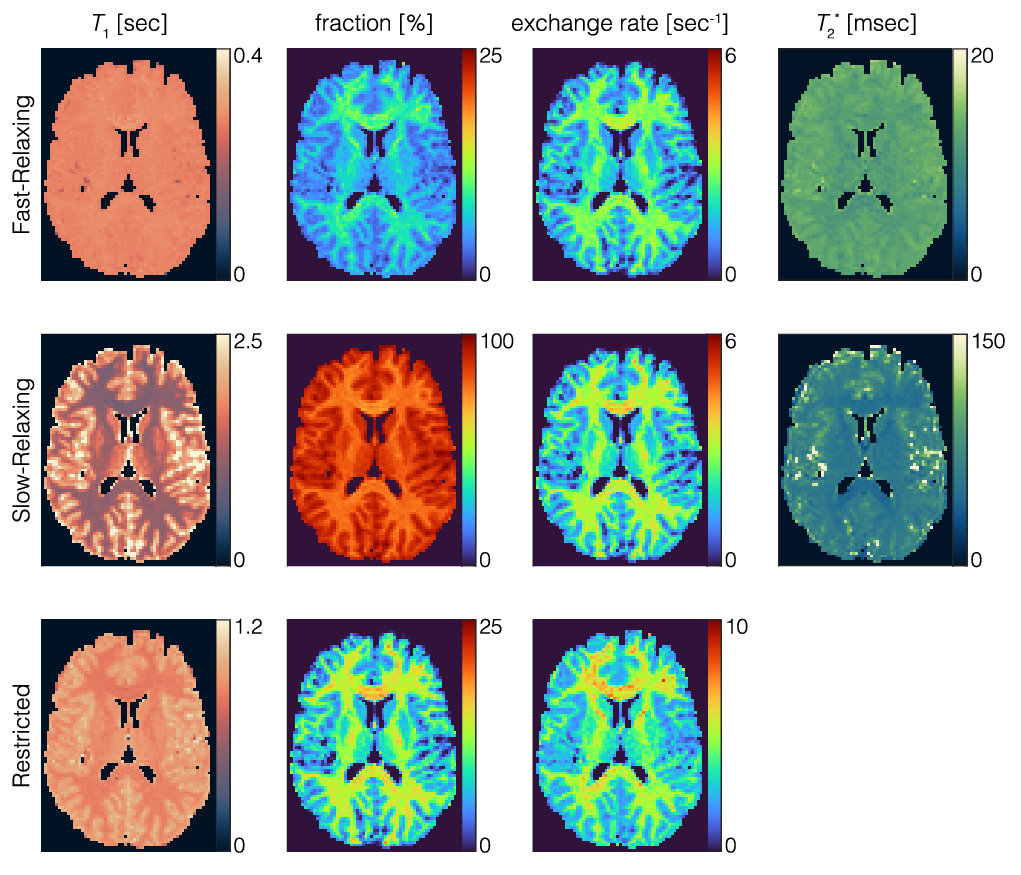}
  \caption{Estimated $T_1$ (column 1), relative fraction (column 2) and exchange rate (column 3) of the fast-relaxing pool (top row), slow-relaxing pool (middle row) and restricted pool (bottom row), along with $T_2^*$ of the fast and slow-relaxing pools (column 4). The fast-relaxing pool shows relatively homogenous $T_1$ values across both the WM and GM regions. In contrast, the slow-relaxing pool exhibits higher $T_1$ values in GM compared to WM regions, as does the restricted pool but with lower contrast. In WM regions, there is an increased fraction of fast-relaxing and restricted pools compared to GM regions, reflecting the greater myelin content in these regions, whereas there is a decreased slow-relaxing pool fraction. This is accompanied by higher exchange rates from F to R, S to R and S to F in WM compared to GM. $T_2^*$ (column 4) of both the fast-relaxing and slow-relaxing pools displays lower values in WM compared to GM regions.}
  \label{fig5}
\end{figure}

in vivo three pool parameter maps of the fast-relaxing (top row), slow-relaxing (middle row) and restricted pool (bottom row) of the brain are presented in Figure \ref{fig5} with regional analysis results listed in Table \ref{table2}. Comparing the spin-lattice relaxation time values (column 1), the fast-relaxing pool shows relatively homogeneous values across both the WM and GM regions. In contrast, the slow-relaxing pool exhibits higher $T_1$ values in GM compared to WM regions, as does the restricted pool but with lower contrast. The relative fraction (column 2) of both fast-relaxing and restricted pools exhibits higher values in myelin-rich white matter regions compared to grey matter regions, whereas the slow-relaxing pool exhibits lower values. This is accompanied by higher exchange rates from F to R, S to R and S to F (column 3) in WM compared to GM. $T_2^*$ (column 4) of both the fast-relaxing and slow-relaxing pools displays lower values in WM compared to GM regions.

Compared to the two-pool BTS method\cite{Jang2023} in Table \ref{table2}, the macromolecular proton fraction $f_\text{R}$ showed relatively good agreement in both WM and GM regions. The free water $T_1$ obtained from the two-pool method ($T_1^\text{F+S}$) was less than the $T_1$ of the slow-relaxing component ($T_1^\text{S}$) but greater than the fast-relaxing component ($T_1^\text{F}$), except for the corpus callosum genu, where the slow-relaxing component showed similar values. Overall, this reflects the single component free water modeling of the two-pool model, effectively averaging the signal from the multiple water compartments.\newline

Further comparing with measurements obtained from a fixed porcine spinal cord WM ex vivo sample using a four-pool model\cite{Wallstein2025} in Table \ref{table2}, the macromolecular proton fraction $f_\text{R}$ shows higher value compared to brain WM regions. $T_1$ of the slow-relaxing component ($T_1^\text{S}$) shows reasonable agreement. However, $T_1$ of the fast-relaxing component ($T_1^\text{F}$) obtained from the four-pool model shows higher value whereas the restricted macromolecule pool ($T_1^\text{R}$) shows lower value. These discrepancies reflect the difference in modeling of interactions between pools of the four-pool model, where the fast-relaxing free pool interacts with both slow-relaxing free pool and fast-relaxing nonaqueous pool, and the slow-relaxing free pool interacts with both fast-relaxing free pool and restricted slow-relaxing nonaqueous pool, in addition to difference in evaluated tissue.

\begin{table}[ht]
\caption{$^1$}
  \begin{threeparttable}
    \centering
    \resizebox{\textwidth}{!}{\begin{tabular}{lccccccccccccccccccccccccc}
      \hhline{==========================}
      \multirow{3}{*}{ROI} & \multicolumn{4}{c}{BTS (2-pool)$^{36}$} & \multirow{2}{*}{} & \multicolumn{11}{c}{MC BTS (3-pool)} & \multirow{2}{*}{} & \multicolumn{8}{c}{MC BTS (4-pool)} \\ 
      \cline{2-5} \cline{7-17} \cline{19-26}
 & $T_1^\text{F+S}$ & $T_1^\text{R}$ & $f_\text{R}$ & $k_\text{[F+S]R}$ & & $T_1^\text{F}$ & $T_1^\text{S}$ & $T_1^\text{R}$ & $f_\text{F}$ & $f_\text{S}$ & $f_\text{R}$ & $k_\text{FR}$ & $k_\text{SR}$ & $k_\text{SF}$ & $T_2^\text{F*}$ & $T_2^\text{S*}$ & & $T_1^\text{F}$ & $T_1^\text{S}$ & $T_1^\text{R}$ & $f_\text{F}$ & $f_\text{S}$ & $f_\text{R}$ & $k_\text{FR}$ & $k_\text{SF}$ \\
 & [s] & [s] & $[\%]$ & [s$^{-1}$] & & [s] & [s] & [s] & [\%] & [\%] & [\%] & [s$^{-1}$] & [s$^{-1}$] & [s$^{-1}$] & [s] & [s] & & [s] & [s] & [s] & [\%] & [\%] & [\%] & [s$^{-1}$] & [s$^{-1}$] \\
      \hhline{==========================} \\
      WM region & & & & & & & & & & & & & & & & & & & & & & & & &\\
      \ \ Corpus & \multirow{3}{*}{\minitab[c]{0.969\\± 0.22}} & \multirow{3}{*}{1.0$^2$} & \multirow{3}{*}{\minitab[c]{13.2\\± 3.6}} & \multirow{3}{*}{\minitab[c]{3.09\\± 0.86}} & & \multirow{3}{*}{\minitab[c]{0.285\\± 0.010}} & \multirow{3}{*}{\minitab[c]{0.96\\± 0.21}} & \multirow{3}{*}{\minitab[c]{0.79\\± 0.06}} & \multirow{3}{*}{\minitab[c]{13.2\\± 5.1}} &\multirow{3}{*}{\minitab[c]{70.8\\± 6.0}} & \multirow{3}{*}{\minitab[c]{16.0\\± 3.1}} & \multirow{3}{*}{\minitab[c]{2.69\\± 1.10}} & \multirow{3}{*}{\minitab[c]{3.69\\± 0.72}} & \multirow{3}{*}{\minitab[c]{7.14\\± 3.47}} & \multirow{3}{*}{\minitab[c]{12\\± 1}} & \multirow{3}{*}{\minitab[c]{57\\± 13}} & & & & & & & & &\\
      \ \ callosum, & & & & & & & & & & & & & & & & & & & & & & & & &\\
      \ \ genu & & & & & & & & & & & & & & & & & & & & & & & & &\\
      \ \ Corpus & \multirow{3}{*}{\minitab[c]{1.020\\± 0.26}} & \multirow{3}{*}{1.0$^2$} & \multirow{3}{*}{\minitab[c]{12.2\\± 3.3}} & \multirow{3}{*}{\minitab[c]{2.88\\± 0.8}} & & \multirow{3}{*}{\minitab[c]{0.283\\± 0.006}} & \multirow{3}{*}{\minitab[c]{1.34\\± 0.33}} & \multirow{3}{*}{\minitab[c]{0.87\\± 0.07}} & \multirow{3}{*}{\minitab[c]{9.0\\± 3.5}} & \multirow{3}{*}{\minitab[c]{77.6\\± 5.5}} & \multirow{3}{*}{\minitab[c]{13.4\\± 2.8}} & \multirow{3}{*}{\minitab[c]{2.15\\± 0.92}} & \multirow{3}{*}{\minitab[c]{3.08\\± 0.65}} & \multirow{3}{*}{\minitab[c]{4.65\\± 2.07}} & \multirow{3}{*}{\minitab[c]{11\\± 1}} & \multirow{3}{*}{\minitab[c]{51\\± 8}} & & & & & & & & &\\
      \ \ callosum, & & & & & & & & & & & & & & & & & & & & & & & & &\\
      \ \ splenium & & & & & & & & & & & & & & & & & & & & & & & & &\\
      \multirow{3}{*}{\minitab[l]{Frontal white,\\matter}} & \multirow{3}{*}{\minitab[c]{0.864\\± 0.18}} & \multirow{3}{*}{1.0$^2$} & \multirow{3}{*}{\minitab[c]{13.2\\± 2.1}} & \multirow{3}{*}{\minitab[c]{3.10\\± 0.51}} & & \multirow{3}{*}{\minitab[c]{0.282\\± 0.006}} & \multirow{3}{*}{\minitab[c]{1.23\\± 0.39}} & \multirow{3}{*}{\minitab[c]{0.83\\± 0.07}} & \multirow{3}{*}{\minitab[c]{9.5\\± 3.9}} & \multirow{3}{*}{\minitab[c]{78\\± 6.5}} & \multirow{3}{*}{\minitab[c]{12.5\\± 3.2}} & \multirow{3}{*}{\minitab[c]{2.12\\± 0.94}} & \multirow{3}{*}{\minitab[c]{2.88\\± 0.73}} & \multirow{3}{*}{\minitab[c]{5.12\\± 2.48}} & \multirow{3}{*}{\minitab[c]{12\\± 1}} & \multirow{3}{*}{\minitab[c]{63\\± 13}} & & \multirow{3}{*}{\minitab[c]{0.714\\± 0.01}} & \multirow{3}{*}{\minitab[c]{1.587\\± 0.02}} & \multirow{3}{*}{\minitab[c]{0.284\\± 0.01}} & \multirow{3}{*}{\minitab[c]{43.4\\± 1.3}} & \multirow{3}{*}{\minitab[c]{33.9\\± 1.1}} & \multirow{3}{*}{\minitab[c]{18.1\\± 0.3}} & \multirow{3}{*}{\minitab[c]{36.5\\± 0.5}} & \multirow{3}{*}{\minitab[c]{1.1\\± 0.1}} \\
%\ \ matter & & & & & & & & & & & & & & & & & & & & & & & & &\\
       & & & & & & & & & & & & & & & & & & & & & & & & &\\
       & & & & & & & & & & & & & & & & & & & & & & & & &\\
      GM region & & & & & & & & & & & & & & & & & & & & & & & & &\\
      \multirow{3}{*}{\minitab[l]{Caudate\\nucleus}} & \multirow{3}{*}{\minitab[c]{1.317\\± 0.18}} & \multirow{3}{*}{1.0$^2$} & \multirow{3}{*}{\minitab[c]{6.7\\± 1.2}} & \multirow{3}{*}{\minitab[c]{1.55\\± 0.29}} & & \multirow{3}{*}{\minitab[c]{0.278\\± 0.004}} & \multirow{3}{*}{\minitab[c]{1.60\\± 0.36}} & \multirow{3}{*}{\minitab[c]{0.86\\± 0.07}} & \multirow{3}{*}{\minitab[c]{6.7\\± 2.4}} & \multirow{3}{*}{\minitab[c]{85.9\\± 4.2}} & \multirow{3}{*}{\minitab[c]{7.5\\± 2.3}} & \multirow{3}{*}{\minitab[c]{1.34\\± 0.59}} & \multirow{3}{*}{\minitab[c]{1.72\\± 0.53}} & \multirow{3}{*}{\minitab[c]{3.69\\± 1.54}} & \multirow{3}{*}{\minitab[c]{13\\± 1}} & \multirow{3}{*}{\minitab[c]{61\\± 13}} & & & & & & & & &\\
       & & & & & & & & & & & & & & & & & & & & & & & & &\\
       & & & & & & & & & & & & & & & & & & & & & & & & &\\
      \multirow{3}{*}{\minitab[l]{Cerebral\\cortex}} & \multirow{3}{*}{\minitab[c]{1.408\\± 0.18}} & \multirow{3}{*}{1.0$^2$} & \multirow{3}{*}{\minitab[c]{6.5\\± 2.2}} & \multirow{3}{*}{\minitab[c]{1.51\\± 0.51}} & & \multirow{3}{*}{\minitab[c]{0.274\\± 0.009}} & \multirow{3}{*}{\minitab[c]{1.65\\± 0.53}} & \multirow{3}{*}{\minitab[c]{0.88\\± 0.09}} & \multirow{3}{*}{\minitab[c]{6.5\\± 3.5}} & \multirow{3}{*}{\minitab[c]{86.3\\± 6.1}} & \multirow{3}{*}{\minitab[c]{7.2\\± 3.6}} & \multirow{3}{*}{\minitab[c]{1.19\\± 0.81}} & \multirow{3}{*}{\minitab[c]{1.66\\± 0.83}} & \multirow{3}{*}{\minitab[c]{3.46\\± 2.24}} & \multirow{3}{*}{\minitab[c]{13\\± 1}} & \multirow{3}{*}{\minitab[c]{83\\± 24}} & & & & & & & & &\\
       & & & & & & & & & & & & & & & & & & & & & & & & &\\
       & & & & & & & & & & & & & & & & & & & & & & & & &\\
      \multirow{3}{*}{\ \ Thalamus} & \multirow{3}{*}{\minitab[c]{1.319\\± 0.43}} & \multirow{3}{*}{1.0$^2$} & \multirow{3}{*}{\minitab[c]{8.6\\± 2.9}} & \multirow{3}{*}{\minitab[c]{2.00\\± 0.69}} & & \multirow{3}{*}{\minitab[c]{0.277\\± 0.004}} & \multirow{3}{*}{\minitab[c]{1.72\\± 0.48}} & \multirow{3}{*}{\minitab[c]{0.91\\± 0.10}} & \multirow{3}{*}{\minitab[c]{6.2\\± 2.8}} & \multirow{3}{*}{\minitab[c]{85\\± 4.4}} & \multirow{3}{*}{\minitab[c]{8.8\\± 2.1}} & \multirow{3}{*}{\minitab[c]{1.75\\± 0.42}} & \multirow{3}{*}{\minitab[c]{2.01\\± 0.49}} & \multirow{3}{*}{\minitab[c]{3.75\\± 1.67}} & \multirow{3}{*}{\minitab[c]{12\\± 1}} & \multirow{3}{*}{\minitab[c]{56\\± 10}} & & & & & & & & &\\
      \\ \\
    \end{tabular}}
      \begin{tablenotes}\tiny
      \item[1] Data acquired are presented as mean ± SD.
      \item[2] $T_1^\text{R}$ = 1 [s] was used in the 2-pool BTS.
      \item[3] 4-pool model parameters obtained from fixed porcine spinal cord WM ex vivo sample63.
    \end{tablenotes}
  \end{threeparttable}
  \label{table2}
\end{table}

\subsubsection{In Vivo Knee}
\begin{figure}
  \centering
  \includegraphics{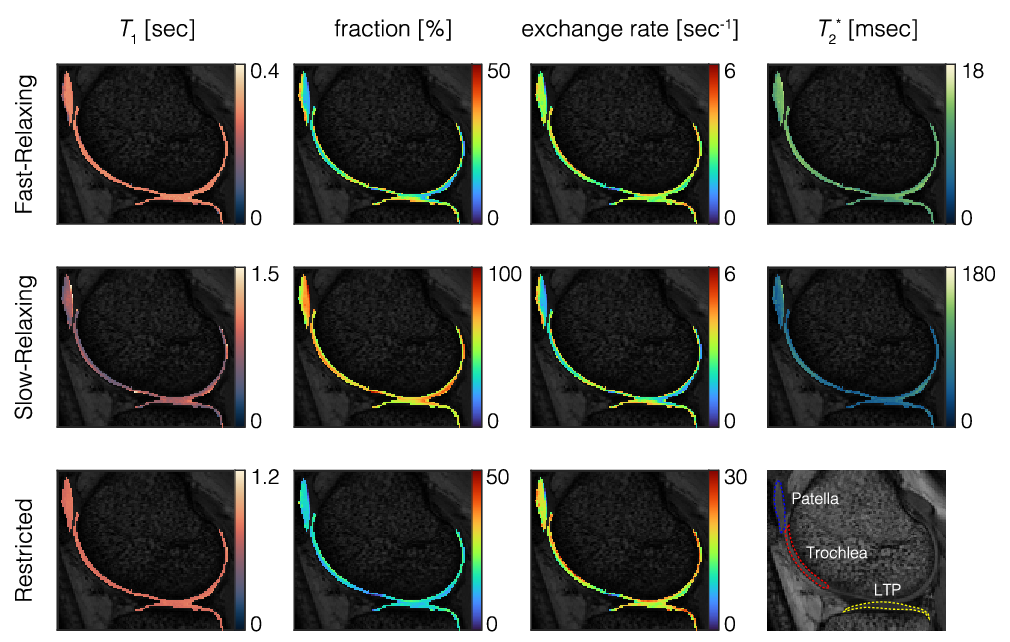}
  \caption{Estimated $T_1$ (column 1), relative fraction (column 2) and exchange rate (column 3) of the fast-relaxing pool (top row), slow-relaxing pool (middle row) and restricted pool (bottom row), along with $T_2^*$ of the fast and slow-relaxing pools (column 4). Comparing $T_1$ values, fast-relaxing and restricted pools show relatively homogenous values across the Patella, Trochlea and Lateral Tibial Plateau (LTP). In contrast, the slow-relaxing pool exhibits higher $T_1$ values in the Patella and LTP compared to the Trochlea. The relative fraction of all three pools exhibits a depth dependence between the deep and superficial layers of the cartilage, where for the fast-relaxing and restricted pool, it is lower in the superficial layer compared to the deep layer, and vice versa for the slow-relaxing pool. This is accompanied by depth-dependent exchange rates. $T_2^*$ of both the fast-relaxing and slow-relaxing pools exhibits depth dependence with slightly higher values in the more superficial layer.}
  \label{fig6}
\end{figure}

in vivo three pool parameter maps of the fast-relaxing (top row), slow-relaxing (middle row) and restricted pool (bottom row) of the knee cartilage are presented in Figure \ref{fig6} with regional analysis results listed in Table \ref{table3}. Comparing the spin-lattice relaxation time values (column 1), the fast-relaxing pool and restricted pool show relatively homogenous values across the Patella, Trochlea and Lateral Tibial Plateau (LTP). In contrast, the slow-relaxing pool exhibits higher $T_1$ values in the Patella and LTP compared to the Trochlea. The relative fraction (column 2) of all three pools exhibits a depth dependence between the deep and superficial layers of the cartilage, where for the fast-relaxing and restricted pool, it is lower in the more superficial layer compared to the deep layer, and vice versa for the slow-relaxing pool. This is accompanied by depth-dependent exchange rates (column 3). $T_2^*$ (column 4) of both the fast-relaxing and slow-relaxing pools exhibits a depth dependence with slightly higher values in the more superficial layer. 

\begin{table}[ht]
\caption{$^1$}
  \begin{threeparttable}
    \centering
    \resizebox{\textwidth}{!}{\begin{tabular}{lccccccccccccccccccccccccc}
      \hhline{==========================}
      \multirow{3}{*}{ROI} & \multicolumn{11}{c}{MC BTS (3-pool)} & \multirow{2}{*}{} & \multicolumn{5}{c}{Liu et al.$^2$} \\ 
      \cline{2-12} \cline{14-18}
       & $T_1^\text{F}$ & $T_1^\text{S}$ & $T_1^\text{R}$ & $f_\text{F}$ & $f_\text{S}$ & $f_\text{R}$ & $k_\text{FR}$ & $k_\text{SR}$ & $k_\text{SF}$ & $T_2^\text{F*}$ & $T_2^\text{S*}$ & & $T_1^\text{F}$ & $T_1^\text{S}$ & $T_1^\text{R}$ & $f_\text{F}$ & $f_\text{R}$\\
       & [s] & [s] & [s] & [\%] & [\%] & [\%] & [s$^{-1}$] &  [s$^{-1}$] & [s$^{-1}$] & [s] & [s] & & [s] & [s] & [s] & [\%] & [\%] \\
      \hhline{==========================} \\
      \multirow{2}{*}{\ \ Patella} & \multirow{2}{*}{\minitab[c]{0.280\\± 0.009}} & \multirow{2}{*}{\minitab[c]{0.900\\± 0.246}} & \multirow{2}{*}{\minitab[c]{0.768\\± 0.025}} & \multirow{2}{*}{\minitab[c]{17.5\\± 6.1}} & \multirow{2}{*}{\minitab[c]{69.0\\± 8.9}} & \multirow{2}{*}{\minitab[c]{13.5\\± 4.2}} & \multirow{2}{*}{\minitab[c]{2.70\\± 0.85}} & \multirow{2}{*}{\minitab[c]{3.09\\± 0.97}} & \multirow{2}{*}{\minitab[c]{10.53\\± 3.65}} & \multirow{2}{*}{\minitab[c]{12\\± 1}} & \multirow{2}{*}{\minitab[c]{63\\± 18}} & & \multirow{2}{*}{\minitab[c]{0.412\\± 0.007}} & \multirow{2}{*}{\minitab[c]{1.853\\± 0.055}} & \multirow{2}{*}{1.0$^3$} & \multirow{2}{*}{\minitab[c]{23.5\\± 1.2}} & \multirow{2}{*}{\minitab[c]{12.3\\± 1.2}} \\
      \\ \\
      \multirow{2}{*}{\ \ Trochlea} & \multirow{2}{*}{\minitab[c]{0.277\\± 0.011}} & \multirow{2}{*}{\minitab[c]{0.813\\± 0.154}} & \multirow{2}{*}{\minitab[c]{0.762\\± 0.025}} & \multirow{2}{*}{\minitab[c]{17.9\\± 5.9}} & \multirow{2}{*}{\minitab[c]{66.0\\± 8.0}} & \multirow{2}{*}{\minitab[c]{16.1\\± 4.3}} & \multirow{2}{*}{\minitab[c]{3.21\\± 0.86}} & \multirow{2}{*}{\minitab[c]{3.68\\± 0.99}} & \multirow{2}{*}{\minitab[c]{10.77\\± 3.57}} & \multirow{2}{*}{\minitab[c]{13\\± 1}} & \multirow{2}{*}{\minitab[c]{69\\± 19}} & & \multirow{2}{*}{\minitab[c]{0.401\\± 0.032}} & \multirow{2}{*}{\minitab[c]{1.690\\± 0.089}} & \multirow{2}{*}{1.0$^3$} & \multirow{2}{*}{\minitab[c]{19.5\\± 0.3}} & \multirow{2}{*}{\minitab[c]{13.3\\± 0.7}} \\
      \\ \\
      \multirow{2}{*}{\minitab[l]{Lateral Tibia\\Plateau}} & \multirow{2}{*}{\minitab[c]{0.283\\± 0.013}} & \multirow{2}{*}{\minitab[c]{0.938\\± 0.197}} & \multirow{2}{*}{\minitab[c]{0.770\\± 0.038}} & \multirow{2}{*}{\minitab[c]{20.7\\± 6.6}} & \multirow{2}{*}{\minitab[c]{62.7\\± 8.1}} & \multirow{2}{*}{\minitab[c]{16.6\\± 4.3}} & \multirow{2}{*}{\minitab[c]{3.32\\± 0.87}} & \multirow{2}{*}{\minitab[c]{3.80\\± 1.00}} & \multirow{2}{*}{\minitab[c]{12.44\\± 3.96}} & \multirow{2}{*}{\minitab[c]{11\\± 1}} & \multirow{2}{*}{\minitab[c]{50\\± 12}} & & \multirow{2}{*}{\minitab[c]{0.417\\± 0.023}} & \multirow{2}{*}{\minitab[c]{1.811\\± 0.087}} & \multirow{2}{*}{1.0$^3$} & \multirow{2}{*}{\minitab[c]{23.2\\± 2.5}} & \multirow{2}{*}{\minitab[c]{18.3\\± 6.0}} \\
      \\ \\
    \end{tabular}}
      \begin{tablenotes}\tiny
        \item[1] Data acquired are presented as mean ± SD.
        \item[2] Values reported from Liu et al.47 at 3T, refer to reference for method details.
        \item[3] $T_1^\text{R}$ = 1 [s] was assumed in Liu et al.
      \end{tablenotes}
    \end{threeparttable}
  \label{table3}
\end{table}

\section{DISCUSSIONS AND CONCLUSION}
We have presented a novel signal acquisition and modeling strategy that accounts for the full spin dynamics among fast-relaxing and slow-relaxing aqueous free water pools, as well as a non-aqueous restricted proton pool, to jointly estimate multi-component exchange and magnetization transfer effects through the signal magnitude, while compensating for relative $B_1^+$ inhomogeneity via the signal phase. Starting from the governing Bloch-McConnell equations, we derived an analytical signal equation, which was verified through simulations. Monte Carlo simulations across a range of SNRs further validated the proposed fitting algorithm. in vivo experiments in both the brain and knee demonstrated that our method is broadly applicable across different anatomical regions.

MC BTS extends both the sequence framework and signal modeling strategy of the original BTS method by incorporating a multi-echo acquisition to extract additional spin dynamics information. Without increasing scan time, MC BTS leverages this additional temporal evolution to apply a more advanced three-pool tissue model, enabling estimation of spin-lattice relaxation times, relative pool sizes and exchange rates among the three pools, along with multi-component $T_2^*$ information. In the original BTS method, the spin-lattice relaxation time of the macromolecular proton pool was assumed to be 1 s, a well-accepted assumption due to its minimal influence on steady-state signal behavior\cite{Sled2001,Yarnykh2002,Helms2009,Liu2016a,Henkelman1993,Tyler2005}. In MC BTS, the spin-lattice relaxation time of the macromolecular pool is explicitly estimated during parameter fitting to minimize estimation bias caused by prior tissue assumptions. Our results in the brain demonstrate differences in estimated values between WM and GM regions, reflecting variations in the composition of the macromolecular pool (myelin-rich vs. non-myelin) and aligning with previous reports of distinct spin-relaxation times\cite{Helms2009,Manning2021,Murali-Manohar2021}. In addition, the knee cartilage results show value variation across different cartilage subregions, reflected by their different macromolecular pool composition and mechanical weight-bearing status\cite{Sritanyaratana2014,Liu2016a}. While MC BTS demonstrated the ability to provide specific information about tissue composition and microstructure in normal brain and knee cartilage, its application to pathological tissues has not yet been explored. Further investigation is warranted to evaluate its sensitivity and specificity for detecting tissue degradation in conditions such as multiple sclerosis in the brain or osteoarthritis in cartilage. Moreover, although MC BTS provides extensive multiparametric information, translating these parameters into meaningful physiological or clinical insights remains challenging and will require future validation to support clinical utility.

\begin{figure}
  \centering
  \includegraphics{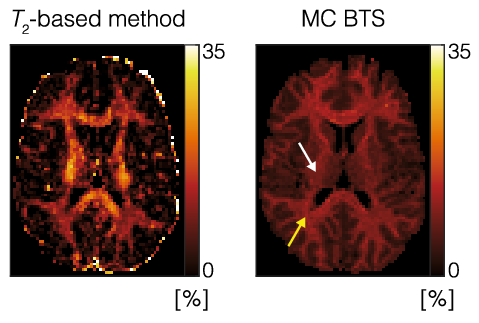}
  \caption{Fast-relaxing free water fraction obtained from the $T_2$-based method (left column) and MC BTS (right column) calculated with respect to free water. Optic radiation and corticospinal tract are indicated by the yellow and white arrows, respectively.}
  \label{fig7}
\end{figure}

Myelin water fraction maps obtained from the $T_2$-based method and MC BTS (i.e., $f_\text{F}$, the fast-relaxing free water fraction) are shown in the left and right columns of Figure \ref{fig7}, respectively. The myelin water fraction values derived from the $T_2$-based method appear relatively inhomogeneous, particularly within WM regions, where the corticospinal tract shows higher myelin water content than its neighboring WM areas. In contrast, the maps from MC BTS are more uniform and show well-differentiated values between WM and GM. This aligns more closely with $T_1$-based myelin water imaging methods\cite{Oh2013,Ma2022}, which leverage the short $T_1$ of myelin water to suppress the long $T_1$ signal of axonal/extracellular water via inversion recovery, thereby improving SNR and reducing artifacts. In the frontal WM and the splenium of the corpus callosum, the myelin water fraction measured with MC BTS was 10.9\% and 10.4\%, respectively, consistent with previous $T_2^*$-based studies performed at 3T\cite{Du2007}. In contrast, the $T_2$-based method produced higher estimates of 13.7\% and 19.4\% in these regions. Although a previous comparison study has shown that $T_2$-based methods tend to estimate higher myelin water fraction values relative to $T_2^*$-based methods\cite{Alonso-Ortiz2018}, the difference observed in our study was more pronounced, likely due to several factors. In earlier work\cite{Alonso-Ortiz2018}, the comparison involved a multi-echo spin-echo sequence and a multi-echo gradient-echo sequence, without corrections for MT effects. By contrast, our study compares a multi-echo gradient-echo MC BTS sequence, which includes MT and $B_1^+$ correction, with a more recent implementation of a 3D GRASE sequence, a modified multi-echo spin-echo approach. Additionally, the GRASE sequence used in our study employed a TR of 1 second, which could introduce partial relaxation effects into the acquired signal. These differences in acquisition protocols and modeling strategies may explain the discrepancies observed. Another factor potentially contributing to these differences is WM fiber orientation relative to $B_0$, which has been shown to bias myelin water fraction estimation in $T_2$-based methods\cite{Birkl2021}. The angle between WM fibers and the main magnetic field also affects $T_2^*$ estimates. This orientation effect is reflected in the MC BTS maps in Figure \ref{fig7} (right column), where regions perpendicular to $B_0$, such as the optic radiation (yellow arrow), show relatively elevated myelin water fraction compared to parallel regions like the corticospinal tract (white arrow), consistent with previous observations\cite{Chan2020}. Further investigation into how these confounding factors influence myelin water fraction estimation is beyond the scope of this study and will be explored in future work.

Given the large number of parameters and the inherent complexity of the three-pool MC BTS model, the high dimensional cost function is further complicated by the presence of local minima and saddle points\cite{Waterfall2006,Transtrum2010,Dauphin2014}, a challenge that is exacerbated by the flatness of least squares residual surfaces. To assess the least squares residual surface structure and noise performance, a minimum least squares residuals (MLSR) analysis\cite{Bouhrara2016a} of the three-pool MC BTS model and two-pool BTS model was carried out (see Supplementary Appendix S1). Comparing common model parameters $f_\text{R}$ and $T_1^\text{S}$, $f_\text{R}$ exhibits similar cost signature whereas $T_1^\text{S}$ of the two-pool BTS shows higher concavity compared to the three-pool MC BTS model, reflecting the intricacy and high number of parameters of the three-pool MC BTS model. Performing sensitivity analysis with regards to sequence parameters and its impact on parameter quantification (see Supplementary Appendix S2), excitation flip angle showed the highest influence in quantification of model parameters. This is likely due to its influence on all steady-state signals used for quantification in contrast to offset frequency, which influence only the BTS acquisitions. In addition, despite the strong alignment between the MC BTS analytical equation and simulation results, a 0.4 $\sim$ 3.1\% degree of discrepancy exists, which effects quantification accuracy as observed by deviations of 4\% for $k_\text{SR}$, 3.7\% for $f_\text{R}$, 3\% for $k_\text{FR}$ and 2\% for $T_2^\text{F}$ of its respective ground truth in the Monte Carlo simulation results, with the remaining parameters being within 1\%. Based on the above, multiple strategies can be employed to improve the quantification accuracy and noise performance of MC BTS. One approach is to utilize methods such as MLSR\cite{Bouhrara2016a,Bouhrara2016} and Cramer-Rao bound (CRB) analysis to carefully select optimal sequence parameters that give rise to favorable residual cost properties leading to higher quantification accuracy and robustness to noise\cite{Lankford2013,Lewis2016,Zhang2022}. Another method is to directly remove noise from the signal by applying denoising techniques\cite{Kang2024}. Further examination of these strategies lies beyond the scope of this study and will be addressed in future work.

MT asymmetry, which originates from the difference in resonance frequency between the free water and macromolecule pools\cite{Hua2007}, may also impact the accuracy of our proposed method. It can be integrated in our method by introducing an additional parameter that reflects this difference in frequency to our analytical signal model. However, further modification of our current protocol is required to include additional acquisition at negative offset frequencies, leading to prolonged scan times. Utilizing acceleration techniques mentioned below may provide an avenue to investigate the impact MT asymmetry has on our method within a feasible scan time and whether the 4000 Hz off-resonance used in this study is sufficiently large to compensate for MT asymmetry. Moreover, due to the analytical signal equation accounting for on-resonance saturation effects, the choice of cut-off frequency used to estimate the Super-Lorentzian absorption lineshape at the asymptotic limit $\Delta \rightarrow 0$ may affect quantification due to the frequency offset from relayed nuclear Overhauser effects being very close to the chemical shift of the macromolecule proton pool\cite{Singh2023} (see Supplementary Appendix S3). The effects that MT asymmetry have on MC BTS require further investigation which will be pursued in forthcoming studies.

An issue that arises with multiparametric methods, including ours, is the long acquisition time. Although we have leveraged the idle time of the relatively long TR stemming from SAR constraints and spoiling requirements to obtain more parametric information at no additional scan time, there is a need to shorten the current acquisition time to make our method clinically viable. However, careful consideration must be taken of the trade-off between scan time reduction and quantitative accuracy. In this study, parallel imaging with two-fold acceleration was applied to accelerate acquisition time for the brain. Conversely, no acceleration was applied to the knee due to SNR constraints. Acceleration techniques such as parallel imaging not only reduce the overall SNR but also introduce spatial nonuniformity, which can degrade quantification accuracy. To improve scanning efficiency, various $k$-space sampling strategies can be explored, including CAIPIRHINIA\cite{Breuer2006}, compressed-sensing-based non-coherent sampling\cite{Lustig2007} and non-Cartesian sampling\cite{Afshari2025}. Other spatial encoding methods have also been proposed\cite{Jang2024}, which may offer improved SNR performance. Model-based reconstruction approaches can also be used and further combined with parallel imaging to decrease acquisition time. Additionally, recent deep learning methods\cite{Bian2024} for qMRI have demonstrated enhanced noise robustness and improved quantitative accuracy, making them promising options for further improving scanning efficiency.

In summary, we introduced a new method, MC BTS, which simultaneously induces Bloch-Siegert shift and magnetization transfer while employing a three-pool model combined with multi-echo acquisition to characterize multi-component tissues containing macromolecules. An analytical signal model was derived and validated through simulations, and a corresponding fitting algorithm was developed and assessed using Monte Carlo simulations across various SNRs. Finally, the in vivo applicability of MC BTS was demonstrated in both the brain and knee.

\section*{ACKNOWLEDGEMENTS}
Research reported in this publication was supported by the National Institute of Biomedical Imaging and Bioengineering under Award Number R21EB031185, National Institute of Arthritis and Musculoskeletal and Skin Diseases under Award Numbers R01AR081344, R01AR079442, R56AR081017, R01AR078877 and National Institute of Neurological Disorders and Stroke under Award Number R01NS125020.

\bibliographystyle{unsrt}
\bibliography{references}  %%% Remove comment to use the external .bib file (using bibtex).


\end{document}